\documentclass[reprint,superscriptaddress,twocolumn,prx]{revtex4}
\usepackage{amsmath,graphicx}
\usepackage{subfigure}
\usepackage[usenames]{color}
\usepackage{ulem}
\usepackage[colorlinks=true,linkcolor=blue,citecolor=red]{hyperref}%

\usepackage[utf8]{inputenc}
\usepackage{amssymb}
\usepackage[english]{babel}
\DeclareUnicodeCharacter{2212}{-}

\begin{document} 

\title {Gate-Assisted Phase Fluctuations in All-Metallic Josephson Junctions}
\author{J. Basset\footnote{julien.basset@universite-paris-saclay.fr}} 
\affiliation{Universit\'e Paris-Saclay, CNRS, Laboratoire  de  Physique  des  Solides,  91405 Orsay,  France} 
\author{O. Stanisavljevi\'c } 
\affiliation{Universit\'e Paris-Saclay, CNRS, Laboratoire  de  Physique  des  Solides,  91405 Orsay,  France}  
\author{M. Kuzmanovi\'c\footnote{Now at QTF Centre of Excellence, Department of Applied Physics,
		Aalto University School of Science,
		P.O. Box 15100, 00076 Aalto, Finland}} 
\affiliation{Universit\'e Paris-Saclay, CNRS, Laboratoire  de  Physique  des  Solides,  91405 Orsay,  France}
\author{J. Gabelli} 
\affiliation{Universit\'e Paris-Saclay, CNRS, Laboratoire  de  Physique  des  Solides,  91405 Orsay,  France}
\author{C.H.L. Quay} 
\affiliation{Universit\'e Paris-Saclay, CNRS, Laboratoire  de  Physique  des  Solides,  91405 Orsay,  France}
\author{J. Est\`eve} 
\affiliation{Universit\'e Paris-Saclay, CNRS, Laboratoire  de  Physique  des  Solides,  91405 Orsay,  France} 
\author{M. Aprili\footnote{marco.aprili@universite-paris-saclay.fr}} 
\affiliation{Universit\'e Paris-Saclay, CNRS, Laboratoire  de  Physique  des  Solides,  91405 Orsay,  France} 
\date{\today}

\begin{abstract} 
The discovery that a gate electrode suppresses the supercurrent in purely metallic systems is missing a complete physical understanding of the mechanisms at play. We here study the origin of this reduction in a Superconductor-Normal metal-Superconductor Josephson junction by performing, on the same device, a detailed investigation of the gate-dependent switching probability together with the local tunnelling spectroscopy of the normal metal. We demonstrate that high energy electrons leaking from the gate trigger the reduction of the critical current which is accompanied by an important broadening of the switching histograms. The switching rates are well described by an activation formula including an additional term accounting for the injection of rare high energy electrons from the gate. The rate of electrons obtained from the fit remarkably coincides with the independently measured leakage current. Concomitantly, a negligible elevation of the local temperature is found by tunnelling spectroscopy which excludes overheating scenarios.
This incompatibility is resolved by the fact that phase dynamics and thermalization effects occur at different time-scales.
\end{abstract}
\maketitle

\section{Introduction}

Understanding the dynamics of quasiparticle excitations in a superconductor and how they interact with the superconducting condensate is important for understanding dissipation mechanisms responsible for decoherence in superconducting devices \cite{Lutchyn2006,Zgirski2011}.
For example, thermally-induced phase slips of the superconducting order parameter or absorption of electromagnetic radiation locally suppress superconductivity, produces quasiparticles and hence decoherence \cite{Golubev2008,Martinis2009,Vepsalainen2020,Cardani2021}.
Whereas the injection of low energy quasiparticles (meV) in mesoscopic superconductors has been extensively studied \cite{Beckmann2004,Cadden2007,Kleine2010,Quay2013}, little is known about the injection of high energy (eV) quasiparticles \cite{Alegria2020}, especially in the limit of few quanta.

Recently, it has been observed that the critical current of a superconducting weak link formed by a mesoscopic nanowire or constriction \cite{DeSimoni2018,Paolucci2018,Paolucci2019,DeSimoni2019,Paolucci2019bis, Puglia2020,Rocci2020,Paolucci2020bis}, can be suppressed by applying a voltage bias to a side gate in the vicinity of the weak link. This all metal-based superconducting device, which at first sight seems to operate like a semiconducting field effect transistor, could be used in quantum technologies as a tunable inductance or switch provided dissipation is demonstrated to be low \cite{Morpurgo1998,Lee2003,Wagner2019}. 
Beyond such practical implications, the understanding of this unexpected effect is attracting a lot of attention. In particular, even though it is still disputed, there is indirect evidence that the leakage current of high energy electrons from the gate plays a major role in reducing the critical currents of mesoscopic constrictions \cite{Golokolenov2020,Alegria2020,Ritter2020,Ritter2021}.

The goal of this article is to investigate the transition regime in which a supercurrent reduction is induced by a small ($\approx 10$~fA), high energy, quasiparticles leakage rate which already demonstrates a sizeable and puzzling reduction of the supercurrent. In particular, we make use of complementary tools (tunnelling spectroscopy \cite{Giaever1960,Pothier1997,Lesueur2008} and switching probability histograms) to
study the gate-induced modulation of the supercurrent in a superconductor-normal metal-superconductor (SNS) Josephson junction. A side gate allows us to attenuate the critical current of the junction in a way similar to previous studies \cite{DeSimoni2018,Paolucci2018,Paolucci2019,DeSimoni2019,Paolucci2019bis, Puglia2020,Rocci2020,Paolucci2020bis, Ritter2020,Golokolenov2020,Alegria2020,Ritter2021}. The SNS junction contains a 
tunnel electrode to perform tunnel spectroscopy of the proximitized region.
We demonstrate that a current leakage from the gate to the normal metal triggers a supercurrent reduction. This leakage current is composed of rare ($\Gamma_L =I_L/e \approx  2 \times 10^4~\rm{s}^{-1}$) and high energy ($\approx 5~eV$) electrons.
It is accompanied by a large broadening of the switching histograms \cite{Puglia2020} while, at the same time, a small elevation ($\approx 48$~mK for $6.8$~eV) of the local time-averaged electron temperature is found by tunnelling spectroscopy, excluding a simple overheating scenario \cite{Alegria2020}.
Instead, each high energy electron reaching the junction generates a large temperature spike. The small spike rate combined with the high probability of premature switching to the resistive state due to the high energy involved in the spikes leads to an enhancement of the switching rates for small currents. These independent events are however too rare to significantly elevate the time-averaged electron temperature seen by tunnel spectroscopy. 
The values of the leakage current obtained from this model match the leakage current measured independently with high resolution confirming the soundness of our model. 
The proposed scenario is similar to the hotspots created by photons acting on superconducting nanowires or Josephson junction based single photon detectors \cite{Engel2013,McCaughan2019,Walsh2021,Marsili2016,Kadin1996, Goltsman2001}.

The article is organized as follows. After presenting the sample fabrication and the measurement setup in section \ref{Sample}, we investigate the static properties of the gating effect in section \ref{statics} where we study critical current reduction, leakage current and perform tunnel spectroscopy. We then analyse the phase dynamics in section \ref{SwitchHistogram} that we compare to a model of independent high energy quasiparticles reaching the weak link. Discussion of the results, including comparison to other experiments is given in section \ref{Discussion}.

\section{Sample fabrication and experimental setup}
\label{Sample}
\begin{figure*}[htbp]
	\begin{center}
		\includegraphics[width=12.5cm]{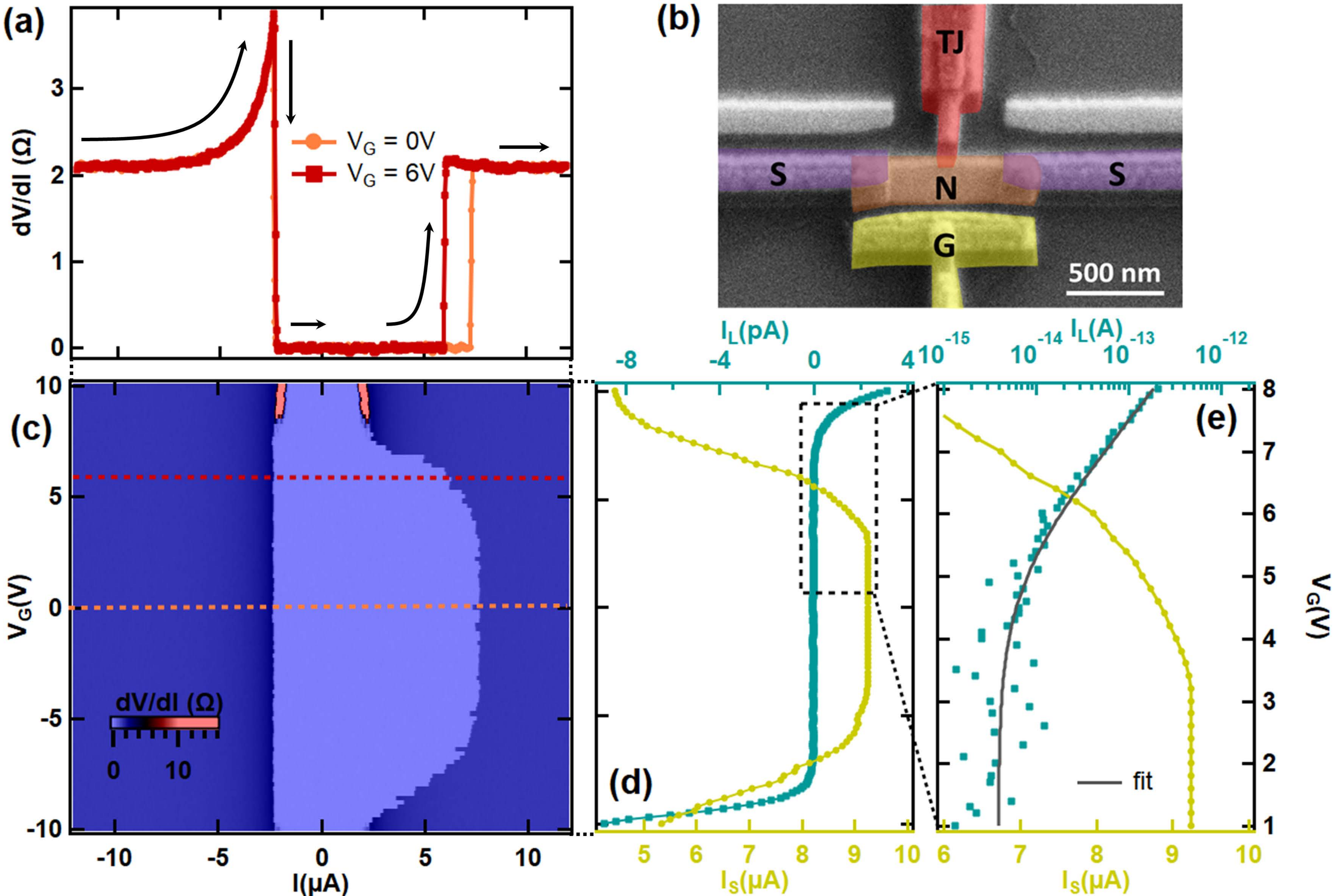}
	\end{center}
	\vspace{-7mm}
	\caption{\textbf{Switching and leakage currents in a gated Josephson junction.} (a) Differential resistance vs current bias $I$ for the two gate voltages $V_G$ indicated by dashed lines in (c). (b) Colorized scanning electron microscope picture of the sample showing the superconducting contacts S, the normal metal N, the tunnel junction electrode TJ and the gate G. (c) Colormap of the differential resistance as a function of  $I$ and $V_G$. (d) Leakage current,$I_L$ and mean switching current $I_S$ as a function of $V_G$. (e) Zoom of (d) in the region of low leakage currents. See supplementary material \cite{SM} for the fit. (a) and (c) share the same x-axis. (c) and (d) share the same y-axis.}
	\label{Figure1}
\end{figure*}
The sample consists of a Superconductor-Normal metal-Superconductor (SNS) Josephson junction tunnel coupled to a superconducting electrode and a side gate (G). All superconducting electrodes are made of Al while the normal metal is Cu. The tunnel junction (TJ) is used to probe the density of states and the local temperature in N \cite{Giaever1960}. A typical sample is shown in figure \ref{Figure1}(b). The tunnel junction surface is roughly $110$~nm$\times 80$~nm and the gate is located $50$~nm away from the SNS junction. The device is fabricated in a single step  e-beam lithography followed by a three angle e-gun evaporation technique. We first deposit a thin $10$~nm layer of Al with a positive angle (red in figure \ref{Figure1}(b)). This layer is then oxidized in pure oxygen to form a good tunnel barrier with $45$~nm of copper evaporated at zero angle  (orange) just after oxidation. A third evaporation, at a negative angle, of $100$~nm of Al defines the superconducting contacts (violet) as well as the gate electrode (yellow) necessary to tune the supercurrent across the SNS junction. The samples are cooled down in a dilution refrigerator at $110$~mK base temperature. Each contacts of the SNS and tunnel junctions are connected through dissipative, highly filtered and closely packed twisted pairs. Of note, the gate electrode is independently connected to a semi-rigid microwave cable to avoid stray leakage currents with the measurement lines (see Supplementary Material for details \cite{SM}). The SNS junction normal state resistance is  $R_n\approx2$~$\Omega$ whereas the tunnel junction resistance is $R_t\approx4$~M$\Omega$. Therefore the tunnel junction is non invasive for voltages lower than $0.5$~mV where the current reaches roughly 100~pA. From the temperature dependence of the tunneling spectroscopy and the geometry, we estimate the Thouless energy $E_{Th}=\hbar D/L^2 =\hbar/\tau_D\approx 14.5~\mu$eV and a diffusion time $\tau_D\approx46$~ps.

\section{Stationary properties}
\label{statics}
\subsection{Gate-dependent supercurrent and leakage current}
\label{Gate}

We first address the gate-induced change of the mean switching current, $I_s$, by measuring the differential resistance $dV/dI (I)$ as a function of the current $I$ and gate voltage $V_G$.  The results are shown  as a colormap in figure \ref{Figure1}(c) with line cuts in figure \ref{Figure1}(a). As in previous reports, no gate-dependence of $I_s$  is observed below a threshold value which in our case is about  $\approx\pm4$~V. For $|V_G|>4$~V and up to $|V_G|\approx 9$~V the switching current is reduced by about  50\%  while the retrapping current, the current at which the voltage drop snaps back to zero, $I_r$ is unaffected. Above $|V_G|\approx 9$~V the hysteresis in the I-V characteristics disappears and $I_s=I_r$. The switching current is suppressed further by increasing $V_G$. We separately measure, with high accuracy \footnote{Such precision was achieved using an automated differential measurement technique between gate on and off as explained in the supplementary materials \cite{SM}}, the leakage current from the gate $I_L$, as a function of the gate voltage (figure \ref{Figure1}(d) and (e)). These data demonstrate  a clear correlation between the critical current reduction and the leakage current.

\subsection{Tunnelling spectroscopy}
\label{tunnel}

\begin{figure*}[htbp]
	\begin{center}
		\includegraphics[width=14.5cm]{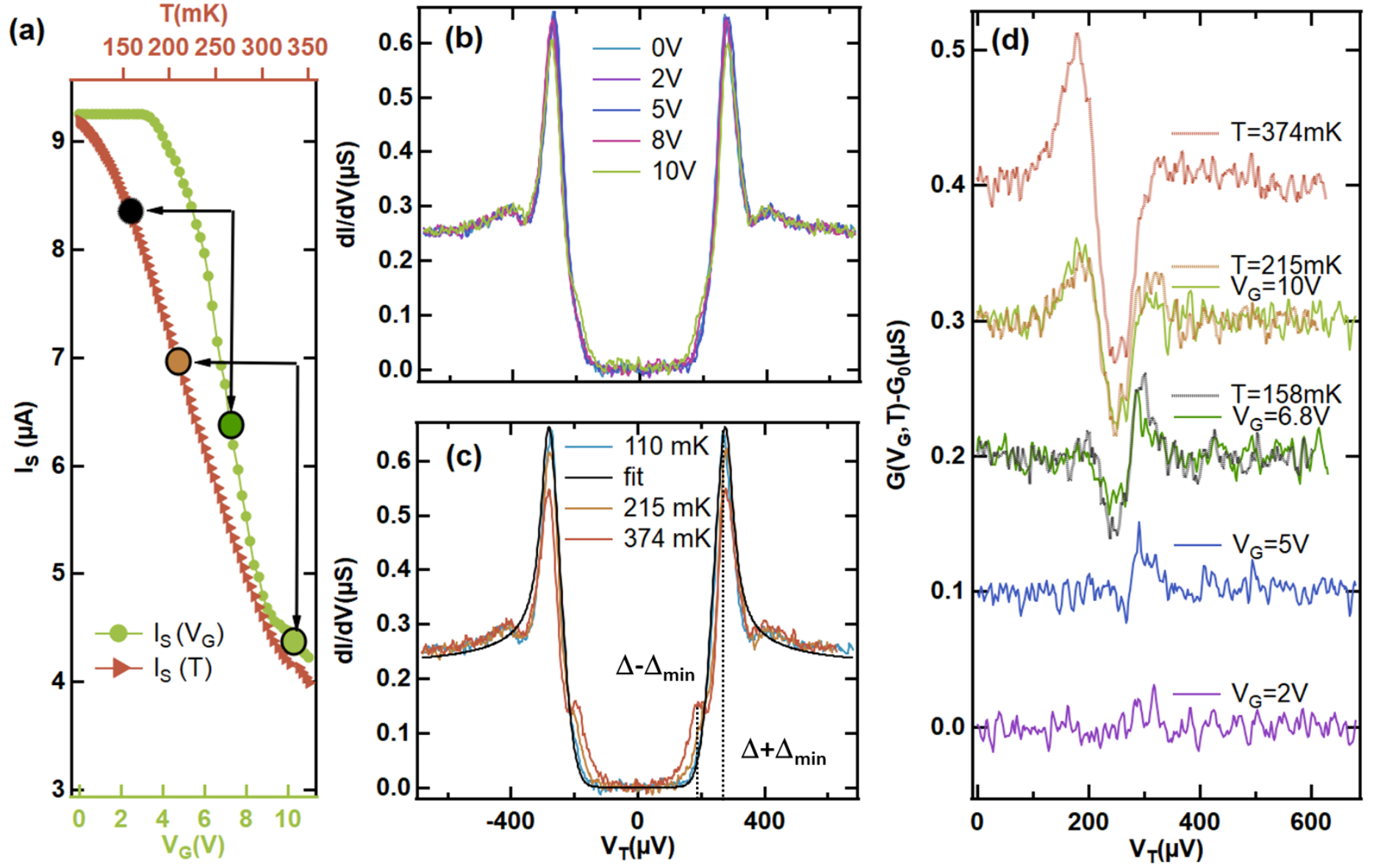}
	\end{center}
	\vspace{-7mm}
	\caption{ \textbf{Local thermometry using tunnelling spectroscopy.} (a) Mean switching current $I_S$ as a function of the bath temperature $T$ and gate voltage $V_G$. (b) Differential conductance of the tunnel junction at different gate voltages and (c) temperatures.  The in-gap feature observed at T$=374$~mK is due to thermally excited quasiparticles through the mini-gap in N. (d) Numerical difference $G(V_G,T)-G_0$ (see text) for different temperatures and gate voltages. The comparison between thermal and gate voltage effects indicates that the decrease in mean switching current can not be accounted quantitatively by overheating. Circles and arrow in (a) show the positions where the electronic temperatures are equivalent (either due to gate or temperature) but not the switching currents.}
	\label{Figure2}
\end{figure*}

In order to show that the reduction of the switching current is not due to overheating, we first measure its temperature and gate dependence $I_S(T)$ and $I_S(V_G)$ shown in figure \ref{Figure2}(a). We then estimate, in a second time, the increase of the electronic temperature $T_e$ due to the gate voltage by tunnelling spectroscopy. The comparison of the temperature and gate datasets allow us to conclude that overheating is, in the limit of small leakage current, a negligible source of reduction of the critical current.

To access experimentally the electronic temperature of the normal metal region, we perform tunnel spectroscopy by measuring the non-linear conductance of the tunnel contact coupled to the N part of the SNS junction (see figure \ref{Figure1}(b)) as a function of gate voltage (figure \ref{Figure2}(b)) and temperature (figure \ref{Figure2}(c)). In this geometry, the quasiparticles current that flows through the junction is given by:
\begin{equation}
I_T(V)=\frac{1}{e R_t}\int dE N_S(E)N_N(E-eV)[f_N(E-eV)-f_S(E)]
\label{eq1}
\end{equation} 
where $N_S$ is the Al BCS density of states and $N_N$ is the density of states in N as modified by the proximity effect. $f_S$ and $f_N$ are the distribution functions.  Assuming that $f_S$ and $f_N$ are the Fermi-Dirac distribution functions at the bath temperature and at $T_e$, respectively, the tunnel current is a direct probe of the time-averaged electron temperature in the normal metal.
According to equation (\ref{eq1}), a rounding of the distribution function, due to an increase of $T_e$ shall be visible for a voltage bias of the tunnel junction  $eV_T\approx\Delta+\Delta_{\rm min}$ where $\Delta=195~\mu$eV is the Al superconducting gap and $\Delta_{\rm min}\approx3.1 E_{\rm Th}\approx45~\mu$eV is the mini-gap induced in N \cite{Lesueur2008}.

%
We show in figure \ref{Figure2}(b) the differential conductance of the tunnel probe as a function of bias voltage $V_T$ for different gate voltages. At the lowest temperature, and only in the extreme case of high gate voltage $V_G=10$~V, a small modification of the differential conductance is visible corresponding to a slight overheating of N.

To estimate $\Delta T$, we compare these gate-dependent datasets to the ones obtained as a function of the bath temperature reported in figure \ref{Figure2}(c). The temperature dependence is well captured by equation (\ref{eq1}). In particular, the high temperature curve exhibits a peak in the differential conductance at $eV_T=\Delta-\Delta_{\rm min}$ when the bath temperature becomes comparable with the minigap $\Delta_{\rm min}$. This peak comes from thermal  quasiparticles in N.
Because the deviations from equilibrium are small we show in figure \ref{Figure2}(d) the numerical difference $G(V_G,T)-G_0$ where $G_0=G(V_G=0~\rm V,T=110~\rm mK)$. This difference shows that gating starts to raise the temperature $T_e$ from $V_G\approx 5$~V. We see experimentally that for $V_G=6.8$~V, the increase in $T_e$ is about 50~mK and 100~mK for $V_G=10$~V as can be seen through the perfect overlaps shown in figure \ref{Figure2}(d).
At the same time, even though the tunneling data coincide, strong differences are observed in the mean switching current (see dots in figure \ref{Figure2}(a)) which is a first hint that the large gate-induced supercurrent reduction observed experimentally cannot be explained by a simple overheating \cite{Alegria2020,Catto2021}. 

We now relate the small raise of temperature to the leakage current. For that we estimate the increase $\Delta T$ in the average electron temperature $T_e$ in the junction due to a flow of high energy electrons. At $V_G =10 $~V where the leakage current reaches $I_L=e \Gamma_L=3~pA$, we get $\Delta T\approx T_0 \tau_R \Gamma_L \approx 100~\rm mK$ with $\tau_R$ the energy relaxation time and  $T_0\approx2$~K the temperature spike generated by a single high energy quasiparticle according to $T_0=(2E/\gamma\Omega)^{0.5}$  where $E\approx eV_G$, $\gamma$ is the Sommerfeld constant of N and $\Omega$ the diffusion volume ($\approx 8\times 10^{-15}$~cm$^3$). The relaxation time in our case is $\tau_R\approx 2.5$~ns, the time for the high energy quasiparticles to escape the same diffusion volume  \footnote{This volume includes both the Copper junction but also the small Al contacts until the larger contacts sink the electrons. The small Al leads need to be taken into account because of the high energy of the electrons which are not bound to Copper}. 
This crude estimation of the temperature rise is comparable to the value obtained by tunnelling spectroscopy.

\section{Switching dynamics}
\label{SwitchHistogram}

\begin{figure*}[htbp]
	\begin{center}
		\includegraphics[width=17cm]{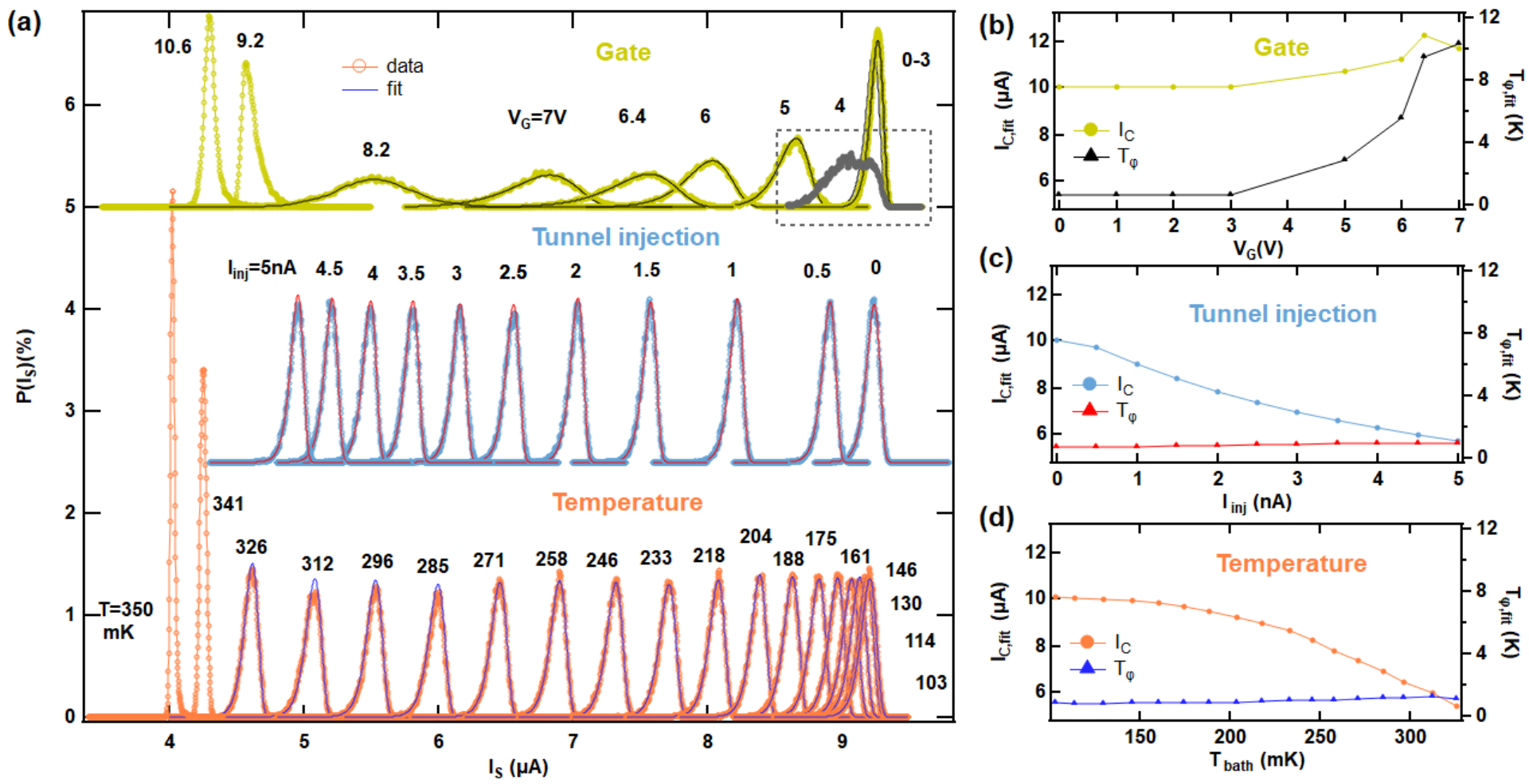}
	\end{center}
	\vspace{-7mm}
	\caption{\textbf{Phase dynamics.} (a) Evolution of the switching probability histograms ($\nu=37$~Hz, $M=12000$) and their fits using equation (\ref{eq3}) with respect to temperature, tunnel current injection and gate voltage. While temperature and tunnel injection show similar behaviour, the gate dependence exhibit a larger broadening of the histograms for the equivalent mean switching current values. At high temperature, the narrowing of the histogram is due to the enhancement of the dissipation. The dashed box highlights the transition regime where equation (\ref{eq3}) cannot explain the data. Evolution of the fitting parameters $I_{C,fit}$ and $T_{\varphi}$  are also shown with respect to (b) gate voltage, (c) tunnel current injection and (d) temperature with the exact same scales.}
	\label{Figure3}
\end{figure*}

We have seen in the previous section that the effect of the gate on the temperature is negligible while the switching current is strongly reduced. This suggests that the phase of the junction is driven out-of-equilibrium by the gate. In order to confirm this out-of-equilibrium scenario, we investigate the switching dynamics.  
To address the gate-induced superconducting phase dynamics \cite{Stewart1968,McCumber1968}, we measure the switching probability histograms for a different set of control parameters (e.g $\nu$, $V_G$, $T$...).  
To do so, we repeatedly (M times) current bias the SNS junction with a sawtooth signal varying from 0 to 10 $\mu$A and back at the frequency $\nu\in[17-277]$~Hz. The junction switches to a resistive state with a certain probability $P(I)$.
We show in figure \ref{Figure3} the histograms obtained for different gate voltages, temperatures and low-energy quasiparticle injection currents. 
The switching probability can be written as a function of the escape rate $\Gamma(I)$ as \cite{Kurkijarvi1972,Fulton1974}:
\begin{equation} 
P(I)=\frac{\Gamma(I)}{\dot{I}}\exp \left[-\int_0^{I} \frac{\Gamma(I')}{\dot{I}}dI'\right]
\label{eq2}
\end{equation} 
where $\dot{I}=dI/dt$ is the speed of the current bias ramp. In the thermally activated regime, the escape rate is given by \cite{Kramers1940,Stewart1968,McCumber1968}:
\begin{equation} 
\Gamma_{\rm{T}}(I)=\frac{\omega_A(I)}{2\pi} \exp \left(-\frac{\Delta U(I)}{k_B T_{\varphi}}\right)
\label{eq3}
\end{equation}
where $\Delta U(I)$ is the potential barrier defined by the tilted-washboard potential, $\omega_A(I)$ is the current dependent plasma frequency and $T_{\varphi}$ is the  phase temperature, which can be higher than the time-averaged electron temperature $T_e$, due to out-of-equilibrium current noise (see Supplementary Materials for more details \cite{SM,Heikkila2002,Basset2019}). 

We first verify that the temperature dependence of the histograms is well captured by the thermal rate expression (\ref{eq3}). The histograms (orange color in figure \ref{Figure3}(a)) slightly broaden and shift to lower current values when the temperature is raised from 100 mK up to roughly 325 mK. Above 330~mK a sharpening of the histograms is observed as the phase dynamics changes from switching to diffusion because of dissipation \cite{Krasnov2005,Mannik2005}. The transition between these two regimes occurs at a temperature corresponding to the mini-gap, as expected, since the number of thermal quasiparticles in N is strongly increased for $k_BT>\Delta_{\rm min}$.
In the restricted range of low temperatures, the histograms are fitted using the thermal rate given in equation (\ref{eq3}).  The critical current $I_C$ and the effective phase temperature $T_{\varphi}$ extracted from the fits are shown in figure \ref{Figure3}(d). $I_c$ diminishes from 10~$\mu$A down to 5~$\mu$A while the phase temperature increases from 800~mK up to 1200~mK. The same analysis can be done for quasiparticles injection from the tunnel junction where a thermal fit reproduces the data using the parameters shown in figure \ref{Figure3}(c). This comparison confirms that injecting low energy quasiparticles ($E<20$~meV)  is equivalent to  raising the average temperature, $T_e$ \footnote{Note that this tunnel current injection corresponds to a voltage ($V_T\in[2,20]$~mV) across the tunnel junction higher than that used for the spectroscopic experiment ($V_T<0.65$~mV) presented above and for which the injected current was too small to perturb the switching mechanism.}.

Comparing the gate dependence to the temperature dependence (figure \ref{Figure3}(a)), the same reduction of the mean switching current by gate voltage leads to much broader histograms \cite{Puglia2020,Ritter2021}. We tentatively fit these histograms using eq.(\ref{eq3}). The quality of the fits is reasonable for most gate voltages but fail to explain histograms close to $V_G=4$~V (see the boxed area at the top of figure \ref{Figure3}(a)) where the supercurrent starts to diminish and the histograms are clearly non-thermal.
Away from this transient regime, the fitting parameters are shown in figure \ref{Figure3}(b). 
The fitted phase temperature rises to 10 K for the largest gate voltage $V_G=10V$ studied here, this is at least 10 times larger than the equilibrium phase temperature and much larger than the electronic temperature observed in tunnelling spectroscopy reported in section \ref{tunnel}. Note also that the increase in $I_c$  obtained from the fits is not physical but effectively compensates the huge increase in the width of histograms.
Finally, we also used the tunnel probe to study the combined action of injecting low energy quasiparticles, equivalent to raising the effective temperature, and high energy electrons leaking from the gate (see Supplementary Material \cite{SM}).
In conclusion, the fitting parameters using eq.(\ref{eq3}) are not physical and confirm that a more sophisticated model than overheating is necessary, especially for the transition regime ($V_G\in[3.2,5.2]$~V) where the histograms cannot be fitted.

To do so, we assume, similarly to single photon detectors \cite{Walsh2021}, that  electrons coming from the gate generates temperature spikes of amplitude $T_S$ equivalent to hot spots \footnote{The effect of the high energy electrons leaking from the gate is modelled by a Poissonian time distribution of the temperature spikes. This reflects the statistics of an independent flow of high energy electrons for small leakage currents. These spikes induce phase fluctuations and do not necessarily switch the junction to the dissipative state if damping is sufficiently high as corroborated by numerical solution of the RCSJ model (see Supplementary Material \cite{SM}).}. The escape rate associated with the gate leakage current is then given by $I_L/e$ multiplied by the escape probability that we choose here to be a thermal escape probability with a temperature $T_S$. With these assumptions, the total escape rate is:
\begin{equation} 
\Gamma(I)= \Gamma_{\rm{T}}(I)+\frac{I_L}{e} \exp \left(-\frac{\Delta U(I)}{k_B T_S}\right).
\label{eq4}
\end{equation}
\begin{figure*}[htbp]
	\begin{center}
		\includegraphics[width=16cm]{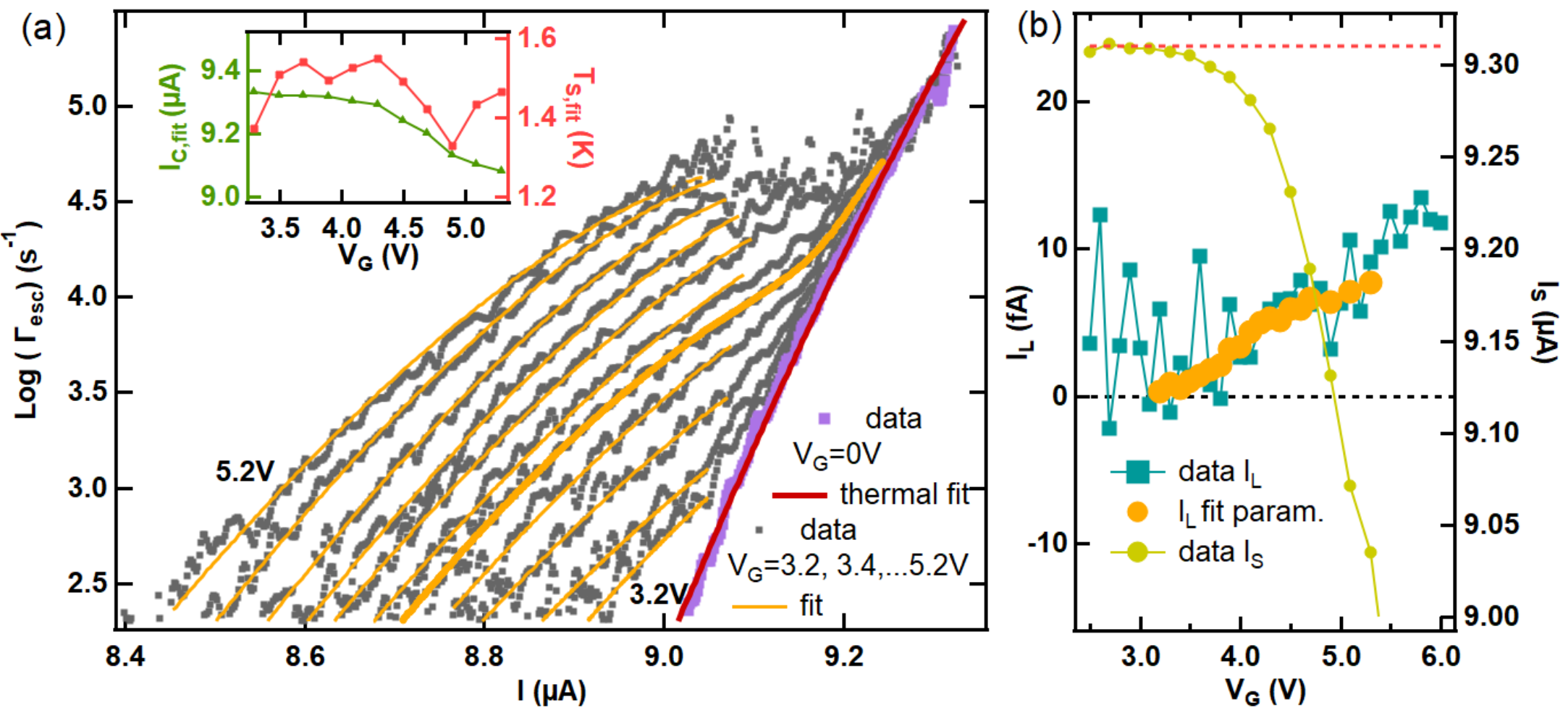}
	\end{center}
	\vspace{-7mm}
	\caption{\textbf{Gate-assisted escape rates.} (a) Escape rates for different $V_G$ as a function of current bias $I$ and the corresponding fits using equation (\ref{eq2}) and (\ref{eq4}). Inset: gate-dependence of the fitting parameters $I_C$ and $T_S$. (b) Directly measured leakage current, mean switching current and the fitting parameter $I_L$ extracted from the fits.}
	\label{Figure4}
\end{figure*}
For a more direct comparison of the data to eq.~(\ref{eq4}) we extract the escape rates as a function of $I$ from the switching histograms inverting eq.~(\ref{eq2}) \footnote{We verified first that the escape rate is independent on the ramp frequency (while the histograms do, see the Supplementary material \cite{SM} for details)}. The escape rates for different gate voltages are shown in Figure \ref{Figure4} in the regime of weak leakage current \footnote{weak leakage currents corresponds to a negligible increase of the time-averaged electron temperature and very likely to a leakage current of uncorrelated electrons as supposed in equation (\ref{eq4})}, $V_G\in[3.2-5.2]$~V.
The log scale graph \ref{Figure4}(a) demonstrates that firstly,  the logarithm of the escape rates are nearly linear in current bias (thus exponential rates), secondly, they grow as the gate voltage/leakage current is raised and thirdly, each curves seems shifted to one another.
Using the second term of eq.(\ref{eq4}), we fit our data with the leakage current $I_L$, the spike temperature $T_S$ and $I_C$ as fitting parameters. 
For $V_G=0$~V (purple dots) the data are well fitted (red line) using only the first term of eq.(\ref{eq4}) and the same parameters ($I_C$ and $T_ {\varphi}$) as for the histograms shown in figure \ref{Figure3}. Increasing $V_G$ requires the second term of eq. (\ref{eq4}) which even becomes dominant. Using both terms in the fitting procedure perfectly reproduces our experimental data (gray dots and orange lines in figure \ref{Figure4}). Importantly, whereas $I_C$ and $T_S$ allow to tackle the full shape of the curve, $I_L$ fixes the shift between the curves.
The values of $I_L$ obtained from the fits as a function of $V_G$ are reported as orange markers in figure \ref{Figure4}(b). They follow well the leakage current measured with a room temperature current amplifier also shown in Figure \ref{Figure4}(b).
The parameters $I_C$ and $T_S$ are shown in the inset of figure \ref{Figure4}(a). Interestingly, we find a quasi gate-independent $T_S$ which is very close from the inferred $T_0=1.5$~K estimated at $V_G=5$~V using the formula proposed at the end of section \ref{statics}.

\section{Discussions and comparison to other experiments}
\label{Discussion}
Researchers from  University of Lancaster \cite{Golokolenov2020} (see also \cite{Catto2021}) recently measured the real and imaginary part of the microwave impedance of a gated Josephson weak link. They show that the field-effect induces large 1/f noise of the microwave impedance. This is coherent with gate-induced non-thermal phase fluctuations. Additionally, the spatially resolved suppression of the critical current reported in reference \cite{Ritter2020} has been attributed to current leakage from the gate. An even more recent work \cite{Ritter2021} emphasizes that the high energy phonons excited by a remote source of high energy electrons alter the supercurrent the same way as a local gate does. This implies that this is the energy brought by the electron that matters and is qualitatively consistent with the temperature spike scenario proposed here.
It is however not clear yet, which is the exact microscopic mechanism at the origin of the leakage current. Some recent studies report that it follows the Fowler-Nordheim model of electron field  emission from  a  metal  electrode \cite{Golokolenov2020,Alegria2020}. In our case (see supplementary material \cite{SM}) but also in \cite{Ritter2021} this does not seem to be the case and it is more likely that the electrons flow either through the substrate or via surface states.   

All these experiments point to out-of-equilibrium effects to explain the gate-induced reduction of the average critical current. 
We have here emphasized that high energy quasiparticles produce large phase fluctuations visible in the broadening of the switching histograms.
Finally, if the leak rate is higher than the relaxation rate the temperature spikes overlap and a global overheating takes place \cite{Alegria2020}.
From our experiment it is not possible to discriminate between charge and energy fluctuations induced by the leakage current, or in other words between the voltage and temperature pulses associated with absorption of hot electrons from the gate. However it is likely that energy imbalance is more relevant than charge imbalance simply because the estimated temperature spike in the normal wire resulting from a 5 eV electron is $T_0 \approx  1.5K$ which sets an energy scale much larger than the chemical potential change that we estimate to $\delta\mu\approx 10^{-8} $ eV. 
 
\section{Conclusion}
\label{Conclusion}

To conclude, we have performed an experiment to probe the field-effect in a metallic Josephson junction by two complementary means, tunnelling spectroscopy and switching histograms. Our experiment reveals that the gate-controllable switching current is triggered by high energy electrons leaking from the gate to the Josephson junction. Their effect is not equivalent to overheating but instead to locally-induced large energy fluctuations which translate in large phase fluctuations visible in the switching histograms. From the escape rates given by the histograms we could extract the leakage current which is in agreement with direct measurements. These findings clarify the local dynamics of high energy quasiparticles and sheds light on the complex mechanisms at play in gated metallic Josephson junctions.

\section*{Acknowledgements}
We thank F. Giazotto, B. Reulet for fruitful discussions and F. Paolucci, F. Giazotto for providing some SNS samples at the beginning of this work. This research has been funded by a ANR JCJC (SPINOES) grant from the Agence Nationale de la Recherche.

\appendix
\renewcommand\thefigure{\thesection.\arabic{figure}}
\setcounter{figure}{0}

\section{Measurement of the gate leakage current}

\begin{figure*}[htbp]  
	\begin{center}
		\includegraphics[width=12.5cm]{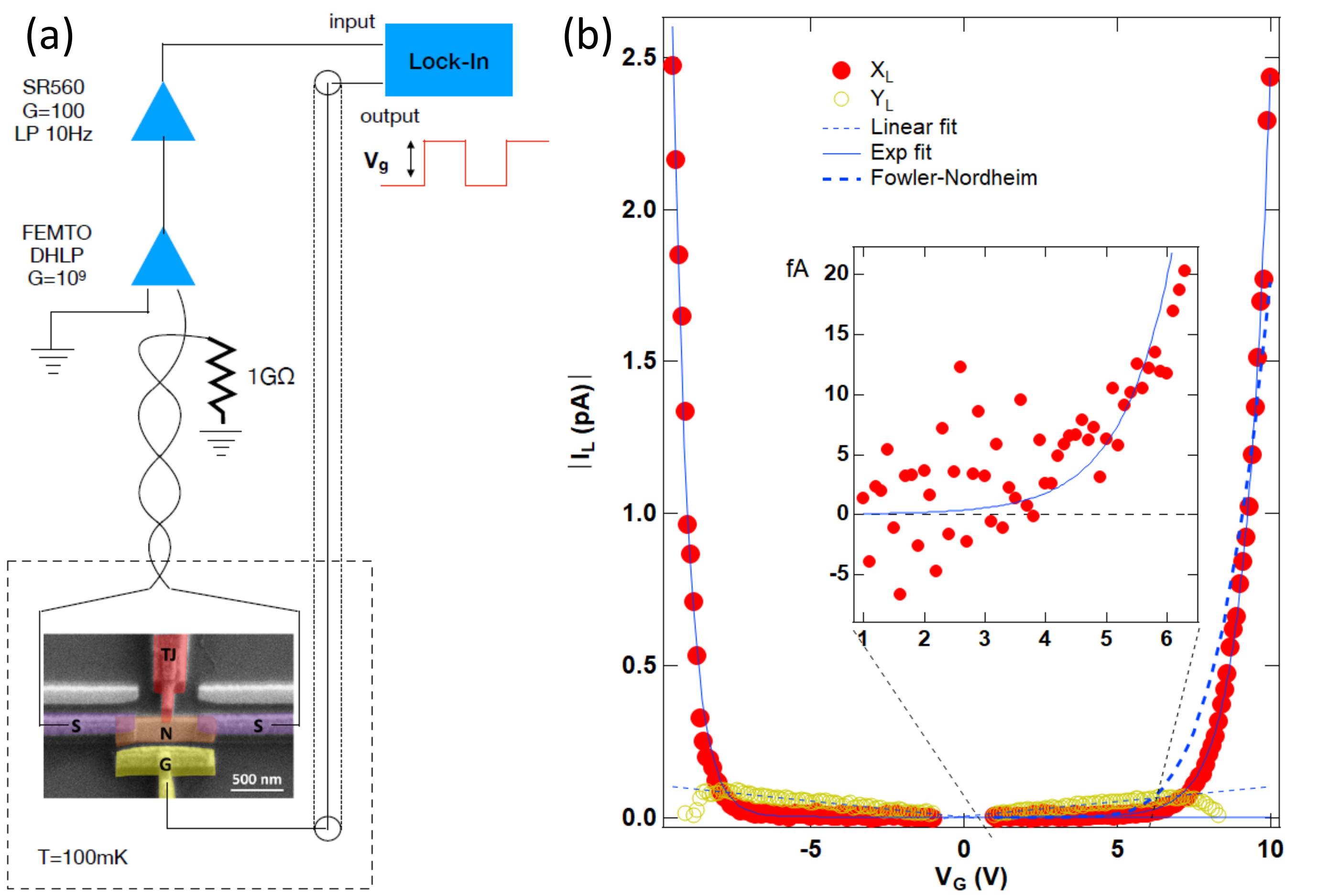}
	\end{center}
	\vspace{-7mm}
	\caption{\textbf{Setup and leakage current }(a) Experimental setup used to precisely measure the leakage current. (b) Absolute values of the in ($\rm{X_L}$) and out ($\rm{Y_L}$) of phase components of the leakage currents measured with the lockin technique (see text for details) and the different fits using equations (\ref{ILND}) and (\ref{ILD}) (see text).}
	\label{FigureS5}
\end{figure*}

The sample is electrically connected to room temperature connectors using twisted resistive wire pairs (the resistance of each wire is about 100 Ohms). To avoid cross-talk and leakage currents, the gate electrode is independently connected using a semi-rigid coaxial cable which is physically separated from the twisted pairs rope.
The leakage current $I_L$,  is measured using the circuit presented in figure \ref{FigureS5}(a). To increase the sensitivity, the gate is biased with a TTL signal of amplitude $V_G$ and frequency equals to 50mHz. $I_L$ is then measured using a lock-in amplifier after amplification and low pass filtering (gain = $10^{11}$ A/V). We verified that a DC measurement of $I_L$ gives the same result with only a worse signal to noise ratio. The absolute values of the in-phase $X_L$, and out-of phase $Y_L$, components of $I_L$ as a function of $V_G$ are presented in figure \ref{FigureS5}(b) for positive and negative values of $V_G$. Note that the values for negative $V_G$ are actually negative. We also verified that a sweep from zero to 10 V in gate voltage gives the same $I_L$ that a sweep back to zero from 10V. $Y_L$ is linear with $V_G$ at low bias as expected for a pure capacitive coupling. The stray capacitance extracted from the fit is C$=10.5 \pm 0.3$~fF. $X_L$ instead is exponential with $V_G$ and corresponds to the dissipative part of the current. The inset of figure \ref{FigureS5}(b) shows a zoom of $X_L$ in the range $1<V_G<6.4$~V. These are the data also presented in figure 4 of the main text.

The microscopic mechanism at the origin of the leakage current remains to be elucidated. Cold emission which has been proposed in \cite{SAlegria2020,SGolokolenov2020} seems unlikely in our case. Indeed, using the Fowler-Nordheim formula \cite{SFowler1928}:
\begin{equation}
I_{L,FN}=\frac{a E^2}{\Phi}e^{\left[-b\Phi^{1.5}/E\right]}
\label{ILND}
\end{equation}
where $a=1.5 \times 10^{-6}$ $A.eV/V^2$,  $b=6.8$~e$V^{-1.5}V/nm$, $\Phi$ is the work function of N (here Cu), $E$ is the electric field, i.e. $V_G/d$ ($d=50nm$) poorly fit the experimental data (see figure \ref{FigureS5}(b)). More importantly the best fitting parameters give a work function for the gate electrode which is more than one order of magnitude smaller than expected. Using the correct work function, the expected leakage current would be negligible. 
A better fit of the data is provided using the function 
\begin{equation}
I_{L,D}=\rm{w}_1\times\rm{exp}\left(w_2(E*d-w_3)\right)
\label{ILD}
\end{equation}
with the coefficients $\rm{w}_1=1.57\times 10^{-16}$~A ($1.24\times 10^{-17}$~A) , $\rm{w}_2= 1.21$~$\rm{V^{-1}}$ (-1.96 $\rm{V^{-1}}$) and $\rm{w}_3= 2$~V (-3.25 V) for positive (negative)  $V_G$, respectively. $I_{L,D}$ is reminiscent of the I-V characteristics of a diode. This supports that the leakage current flows through the Si/SiO2 substrate and is similar to the leakage current between the gate and the source-drain channels in a Metal Oxide Semiconductor Field Effect Transistor \cite{SRalls1984}.

Finally we would like to point out that samples where the distance between the gate and the weak link was about 200 nm did not show any effect on the switching current up to a gate voltage of 120 V.

\section{Numerical solution of the RCSJ model with gaussian (thermal) noise and poissonian (shot) noise}

We describe the phase dynamics within the RCSJ model \cite{SStewart1968,SMcCumber1968} in which the Josephson junction is considered as a parallel circuit of three separated components: a resistance $R$ that accounts for the dissipative current carried in the weak link by the quasiparticles, a capacitance $C$,  and a non-linear inductance due to the finite Josephson current. Following the Kirchhoff's law, the bias current is divided through these three independent elements. Current conservation translates in an equation for the phase dynamics.  Fluctuations are typically accounted by adding a Langevin term that corresponds physically to the thermal noise associated to dissipation in the junction, i.e. the thermal noise of $R$ which is a Gaussian noise. The effect of rare high energy electrons leaking from the gate can be included as a Poissonian noise of Dirac current peaks whose amplitude is larger than $I_c$. This is equivalent to temperature spikes $T_s$ larger than  $T_c$, caused by the absorption of high energy electrons. We here implicitly makes the assumption that energy relaxation is much faster than the average time between two successive events. The stochastic equation governing the phase dynamics can be written as: 
\begin{equation}
\ddot\varphi + \beta \omega_p \dot{\varphi}+\omega_p^2 sin(\varphi)=\omega_p^2 \left[i_b+\delta i+ \sum_{n=1}^{\infty}i_L\delta(t-t_j)\right]
\label{eq1SI}
\end{equation}
where the leakage current is included in the last term $I_L= e/t\sum_{n=1}^{t}\delta(t-t_j)$. We have  introduced the dimensionless bias $i_b=I/I_c$, the Gaussian Langevin term  $\delta i=\delta I/I_c$, the damping parameter $\beta=(RC\omega_p)^{-1}$ and the plasma frequency $\omega_p=\sqrt{2eI_C/\hbar C}$. 

We numerically solve eq.(\ref{eq1SI}) using a Runge-Kutta algorithm \cite{SRunge1895,SKutta1901}. To construct switching histograms we use a linear ramp for the current bias and repeat the ramp $10^4$ times to get a reasonable statistic. The main technical problem to perform such a numerical investigation is that the plasma frequency is about 8 orders of magnitude larger than the ramp frequency that we used experimentally.  To overcome this difficulty, we increase the ramp frequency to  $\omega_p /1000$ and start the ramp sufficiently close from the switching current, $I_s$ to optimize the calculation at currents where the switching probability is sufficiently large. While this procedure is not appropriate to fit the measured histograms, it can however gives some specific insights on the evolution of the histograms and escape rates with increasing leakage current.

\begin{figure*}[htbp]
	\begin{center}
		\includegraphics[width=14.5cm]{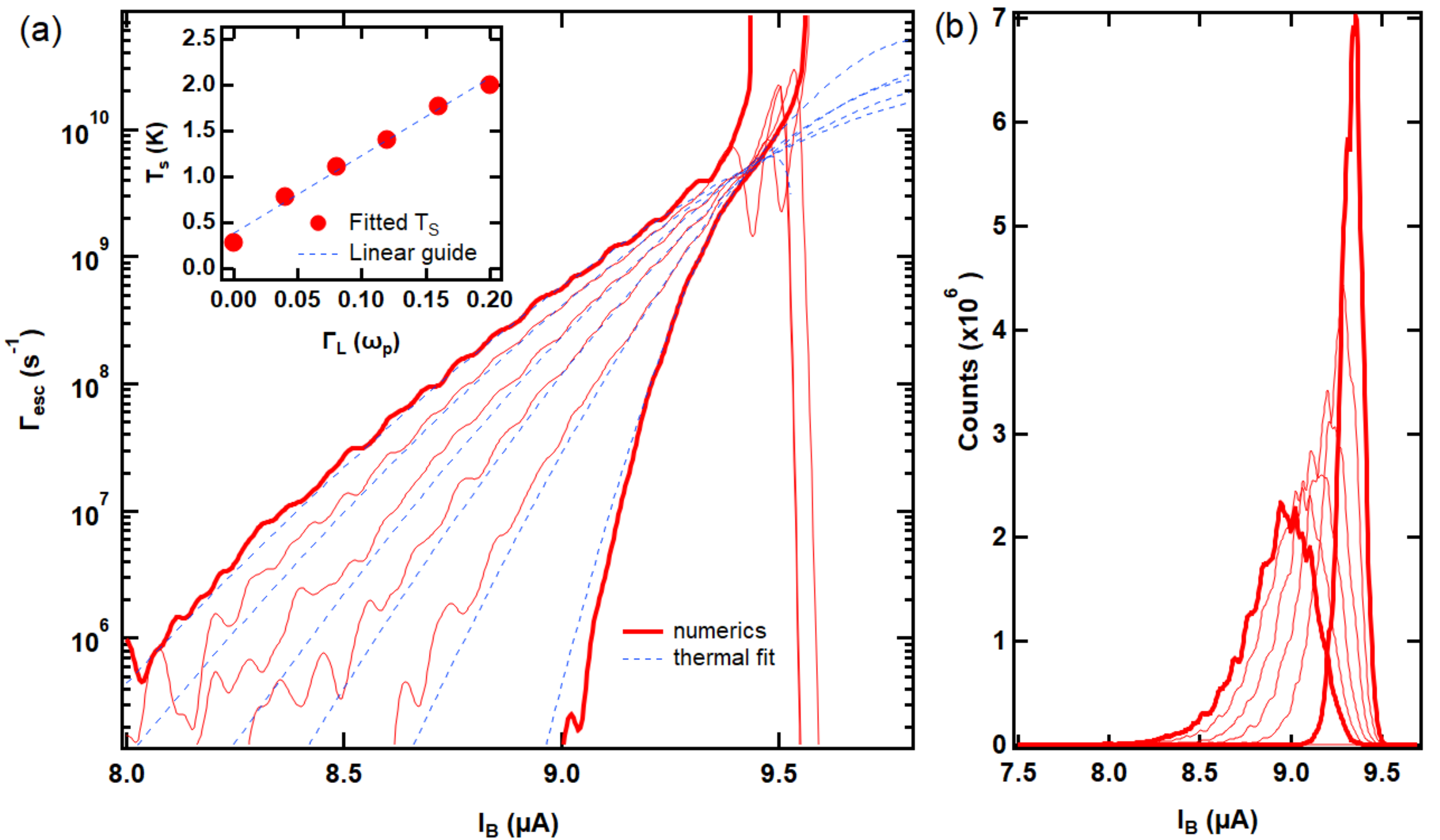}
	\end{center}
	\vspace{-7mm}
	\caption{\textbf{Numerical simulation of switching rates and histograms due to a Poissonian noise source.} (a) Escape rates obtained numerically while varying the electron leakage rate $\Gamma_L/\omega_P$=0, 0.04, 0.08, 0.12, 0.16, 0.2 together with thermal fits. Other parameters are described in the text. Inset: Effective phase temperature vs leakage rate $\Gamma_L$. (b) Corresponding switching histograms.  }
	\label{FigureS1}
\end{figure*}

We first verified that without Poissonian noise, the Gaussian noise reproduces the thermal histograms, then we added the Poissonian noise. The escape rates and switching histograms obtained numerically are presented in figure \ref{FigureS1}(a) and \ref{FigureS1}(b) respectively for $\beta=0.2$ and increasing the leakage current $I_L$ given in units of the plasma frequency multiplied by the electron charge. We remark that when the frequency of the electrons leaking from the gate is sufficiently high, the associated leakage current, whose statistics is Poissonian, leads to a thermal escape rate in which the phase temperature is much higher than the electron temperature in the Josephson junction. This is particularly clear in the inset of figure \ref{FigureS1}(a) where the phase temperature obtained from thermal fits of escape rates is reported as function of  leakage rate. 
Note however that there is a major difference compared to purely thermally-induced switching for which a stronger reduction of the mean value of the switching current is expected. Here instead $I_s$ is only weakly suppressed. This explains the experimental results where the mean value and the variance of the histograms measured as function of the gate voltage and reported in the figure 3 of the main text, are actually uncorrelated. Note also that because of the huge difference in the ramp frequency with respect to the plasma frequency the simulations have been carried out for $I_L$ values of the order of $pA$  for which indeed we recover experimentally a thermal shape of the histograms. Smaller leakage currents are difficult to simulate and the analytical approach described in the main text turns out to be more appropriate.

\section{Switching measurements}

\begin{figure*}[htbp]
	\begin{center}
		\includegraphics[width=14.5cm]{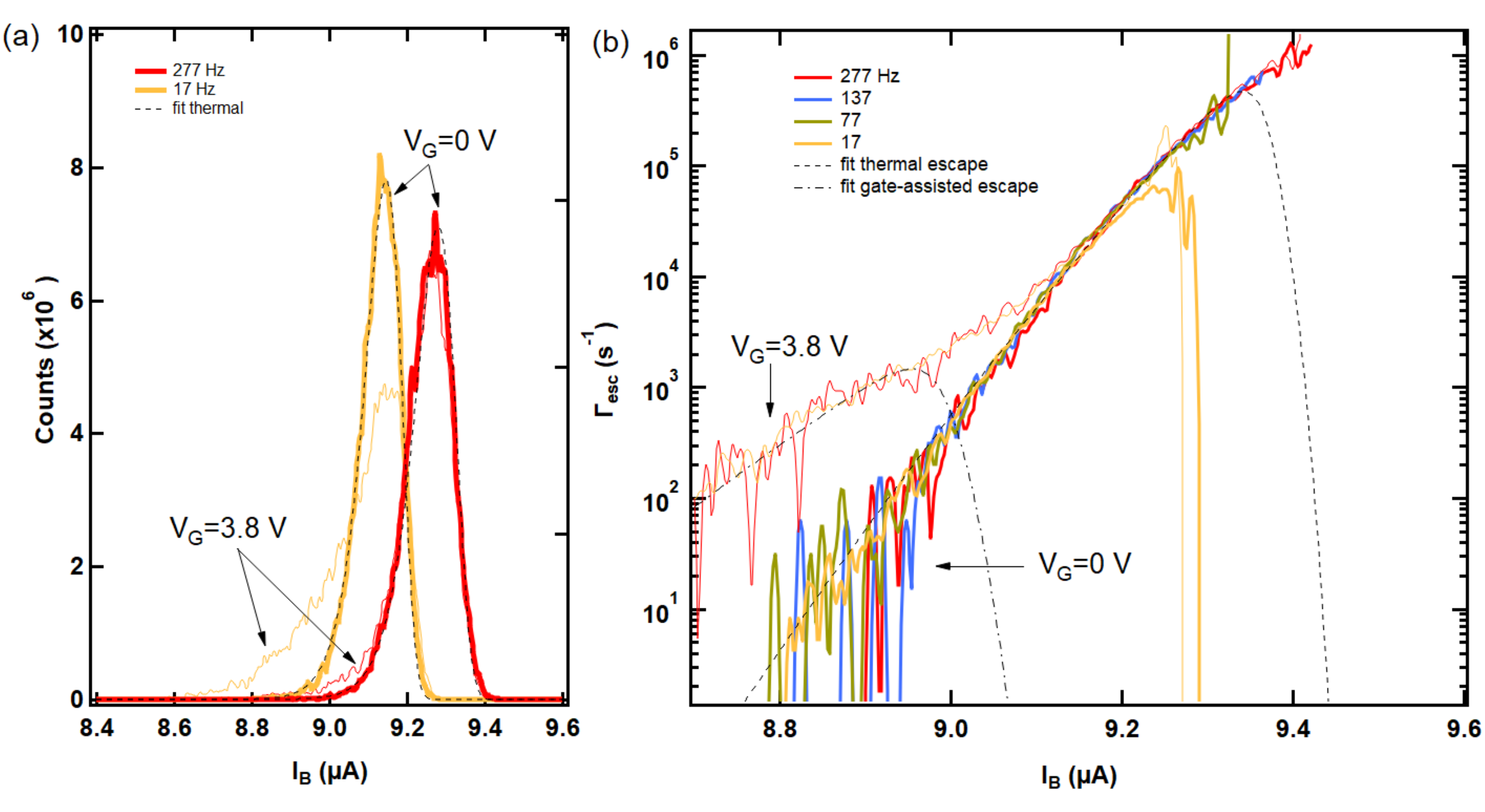}
	\end{center}
	\vspace{-7mm}
	\caption{\textbf{Leakage current-assisted escape rates.} (a) Switching histograms for two frequencies and two gate voltages together with purely thermal fits. Switching histograms depend on the ramp frequency. (b) Escape rates computed from switching histograms obtained at various frequencies. Escape rates are independent of the ramp frequency. At large bias current, thermal fluctuations dominates the escape rate independently of the gate voltage applied. For smaller current bias, the leakage-current assisted escape dominates at finite gate voltage.}
	\label{FigureS3}
\end{figure*}

We report in figure \ref{FigureS3}(a) the switching histograms measured at two different current bias ramp frequencies, $\nu$ = 17 and 277 Hz for $V_G=0$~V (thick line) and $V_G=3.8 V$ (thin line) together with fits using the thermal formula (\ref{eq3}) in the main text \cite{SKurkijarvi1972,SFulton1974} for which the critical current and the phase temperature are fitting parameters. From the $V_G=0$~V fit we get $I_C\approx10~\mu$A and $T_{\varphi}\approx800$~mK. figure \ref{FigureS3}(b) shows the corresponding escape rates obtained from the histograms \cite{SFulton1974}. As expected, while the histograms depend on the speed of the ramp and shift to higher bias current for faster ramps, the rates do not. The reason is that for higher ramp frequencies one explores shorter escape rates as the integrated switching probability is lower for the same bias current, $I$. Note that even in presence of a leakage current induced by a gate voltage, while the histograms are clearly $\nu$ dependent the escape rates are not. To fit the escape rates we use :
\begin{equation} 
\Gamma(I)=\frac{\omega_A(I)}{2\pi} \exp \left(-\frac{\Delta U(I)}{k_B T_{\varphi}}\right)+\frac{I_L}{e} \exp \left(-\frac{\Delta U(I)}{k_B T_s}\right)
\label{eq3SI}
\end{equation}
where $\Delta U(I)$ is the potential barrier defined by the tilted-washboard potential. In the case of long diffusive SNS junctions and low temperature the usual $E_J$cos$(\varphi)$ term has to be replaced by $E_{J} (1-\sum_{n=1}^{\infty}\frac{(-1)^n}{n(2n+1)(2n-1)}\cos n\varphi)$ due to the contribution of higher harmonics to the current-phase relation \cite{SHeikkila2002}.
In the limit $I\rightarrow I_c$, the potential barrier can be approximated by $\Delta U=2E_J(1-I/I_c)^{1.68}$ and the plasma frequency, which defines also the attempt frequency to escape, becomes  $\omega_A(I)=\omega_{p,0}(1-(I /I_c)^2)^{1/4}$ with $\omega_{p,0}=\sqrt{2eI_c/\hbar C}$. The dashed black line in figure \ref{FigureS3}(b) is a fit of the data at $V_G$=0V for which $I_L$=0~fA while the fitting parameters are the same as for the histograms i.e. $T_{\varphi}=0.8 K$ and $I_C=10~\mu$A. The dash-dotted line instead is a fit of the $V_G$=3.8V rates taking into account the second term in eq.(\ref{eq3SI}) as dominant. The corresponding fitting parameters are $I_L=2$~fA and $T_s=1.5K$. These results have been discussed and analysed in the main text.

\section{Combined effect of quasiparticles injection from the tunnel junction and gate}

\begin{figure*}[htbp]
	\begin{center}
		\includegraphics[width=17.0cm]{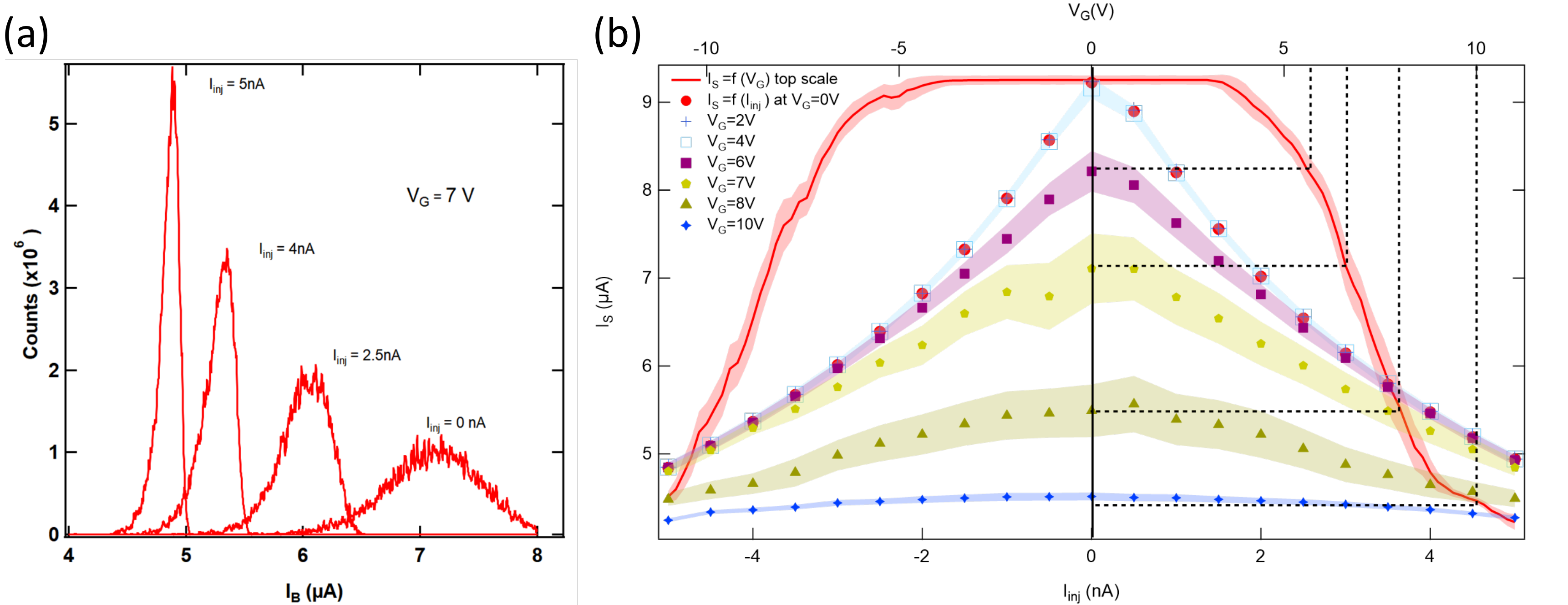}
	\end{center}
	\vspace{-7mm}
	\caption{\textbf{Combined influence of tunnel and gate currents} (a) Switching histograms obtained for $V_G$=7.0V and different injection currents. (b) Top scale: gate-voltage dependence of the mean switching current and its standard deviation represented as shaded error bars. bottom scale: Injection current dependence of the same quantities for different gate voltages. }
	\label{FigureS4}
\end{figure*}

To confirm that the reduction of the switching current, $I_s$, and the broadening of the switching histograms result from the phase dynamics we intentionally increase the damping by injecting low energy quasiparticles using the tunnel junction. While doing so, we make sure that the injection current is always lower than 5 nA and hence more than three orders of magnitude less than $I_s$. As such, it acts as a source of out-of-equilibrium quasiparticles while adding a negligible extra current bias to the weak link.
We show in figure \ref{FigureS4}(a) the switching histograms obtained for $V_G$=7.0V and different injection currents. From these curves, it is obvious that both the mean value and the variance of the distribution are reduced by increasing the amount of injected quasiparticles. A more complete picture is summarized in figure \ref{FigureS4}(b), where we present $I_s$ and the width of the histograms as a function of the injection current $I_{inj}$, for different values of $V_G$. As explained in the main text, the injection of low energy quasiparticles in N is equivalent to raising the electron average temperature. Such elevation has two important effects: it decreases the critical current and the resistance $R$ within the  RCSJ model. A lower $R$ is equivalent to increasing the damping and hence the dissipation as described above. Thus, while the reduction of $I_s$ is a direct consequence of the decrease of $I_c$, the narrowing of the switching histograms is due to the enhanced damping. The shrink of the distribution due to damping finds its origin in the reduction of the attempt frequency in the thermal component of the escape rate. A numerical solution of eq.(\ref{eq1SI}) can also illustrate this result. It is also worth noting that in the case of very high damping the switching distribution becomes a delta function. In the specific case of the leakage current-assisted escape, i.e. the second term of equation (\ref{eq3SI}), the damping does not change the rate of electrons leaking from the gate but reduces the temperature spike, $T_s$, due to absorption of high energy electrons from the gate and hence reduce the energy fluctuations in N.

\end{document}